\documentclass[%
 reprint,
 superscriptaddress,
nofootinbib,
 amsmath,amssymb,
onecolumn
]{revtex4-2}

\usepackage{graphicx}
\usepackage{bm}

\usepackage[
total={6.5in,9.75in}, top=0.9 in, left=0.9in, includefoot,
]{geometry}
\usepackage{color}

\usepackage[normalem]{ulem}

\usepackage{tikz,xcolor,hyperref,textcomp}
\hypersetup{
    colorlinks,
    linkcolor={black!80!black},
    citecolor={black!50!black},
    urlcolor={black!80!black}
}

\definecolor{lime}{HTML}{A6CE39}
\DeclareRobustCommand{\orcidicon}{
	\begin{tikzpicture}
	\draw[lime, fill=lime] (0,0) 
	circle [radius=0.16] 
	node[white] {{\fontfamily{qag}\selectfont \tiny ID}};
	\draw[white, fill=white] (-0.0625,0.095) 
	circle [radius=0.007];
	\end{tikzpicture}
	\hspace{-2mm}
}
	
\foreach \x in {A, ..., Z}{\expandafter\xdef\csname orcid\x\endcsname{\noexpand\href{https://orcid.org/\csname orcidauthor\x\endcsname}
			{\noexpand\orcidicon}}
}

\begin{document}

\preprint{APS/123-QED}

\title{Coexisting chaos and order in micro-textured elastic flows}

 \author{Giulio Foggi Rota\orcidA{}}
 \affiliation{Complex Fluids and Flows Unit, Okinawa Institute of Science and Technology Graduate University, Okinawa 904-0495, Japan}

 \author{Ricardo Arturo Lopez de la Cruz\orcidB{}}
 \affiliation{Micro/Bio/Nanofluidics Unit, Okinawa Institute of Science and Technology Graduate University, Okinawa 904-0495, Japan}
 
 \affiliation{Department of Mechanical Systems Engineering, Faculty of Engineering, Shinshu University, 4-17-1 Wakasato,  Nagano 380-8553, Japan}

 \author{\\ Simon J. Haward\orcidC{}}
 \affiliation{Micro/Bio/Nanofluidics Unit, Okinawa Institute of Science and Technology Graduate University, Okinawa 904-0495, Japan}

 \author{Amy Q. Shen\orcidD{}}
 \affiliation{Micro/Bio/Nanofluidics Unit, Okinawa Institute of Science and Technology Graduate University, Okinawa 904-0495, Japan}

 \author{Marco Edoardo Rosti\orcidE{}}
 \email{marco.rosti@oist.jp}
 \affiliation{Complex Fluids and Flows Unit, Okinawa Institute of Science and Technology Graduate University, Okinawa 904-0495, Japan}

\begin{abstract}
Viscoelastic fluid flows over micro-textured surfaces — densely covered by slender protrusions  like cilia lining the body airways or villi covering the intestinal epithelium — underpin essential biological processes including transport, mixing, and absorption. Despite their ubiquity, the dynamics generated by the interplay between fluid elasticity and these complex geometries remain largely unexplored. Here we combine fully resolved numerical simulations with microfluidic experiments to reveal the flow dynamics established above dense arrays of microscopic pillars (canopies) immersed in the low-Reynolds-number flow of a viscoelastic liquid. We observe that the flow above the canopy tips spontaneously develops elastic turbulence.
Remarkably, the chaotic state coexists with elastic waves emerging from the coupling between fluid elasticity and the heterogeneous shear induced by the canopy geometry. These ordered fluid motions persist across a broad range of flow conditions and canopy configurations.
Our results demonstrate that coherent wave propagation and elastic turbulence are complementary manifestations of viscoelastic fluid flow. Beyond their fundamental significance, these mechanisms have broad implications for transport in biological and engineered environments, and reveal how structured geometries can harness the spontaneous dynamics of viscoelastic liquids to manipulate complex flows.
\end{abstract}

\maketitle

\section{Introduction}

Viscoelastic biological fluids routinely flow over ciliated and villous tissues, where localized shear generated by surface microstructures governs transport, mixing, and physiological function. However, how fluid elasticity interacts with such structured environments remains largely unknown. Unlike Newtonian liquids, viscoelastic fluids store elastic energy that can profoundly alter flow, giving rise to purely elastic instabilities, propagating waves, and chaotic dynamics even when inertia is negligible \cite{spagnolie-2014,bird-1987,page-dubief-kerswell-2020,lellep-linkmann-morozov-2023}. These phenomena are particularly relevant in biological environments, where flows occur at low Reynolds numbers and transport is controlled by the interplay between fluid rheology and tissue morphology.

More generally, flows over arrays of slender surface-mounted elements — commonly referred to as \textit{canopy flows} — occur across a wide range of natural and engineered systems, from aquatic vegetation and atmospheric boundary layers \cite{finnigan-2000,nepf-2012-1} to microfluidic devices and biological epithelia \cite{loiseau-etal-2020, lopez-haward-shen-2025} . Although Newtonian canopy flows have been extensively characterized, particularly in high-Reynolds-number geophysical settings \cite{sundin-bagheri-2019,sharma-garciamayoral-2020-1,monti-etal-2020}, remarkably little is known about how elasticity modifies these flows under the low-Reynolds flow conditions characteristic of biological systems. Recent experiments have shown that viscoelastic canopy flows exhibit spontaneous collective oscillations \cite{deblois-haward-shen-2023} together with the onset of elastic turbulence (ET) \cite{lopez-haward-shen-2025}, a chaotic flow state entirely sustained by elastic stresses \cite{groisman-steinberg-2000,groisman-steinberg-2001-2}. However, the physical mechanisms governing these flows remain elusive.

In particular, coherent wave-like motions have recently been identified in shear viscoelastic flows \cite{page-zaki-2014, page-dubief-kerswell-2020,morozov-2022,foggirota-garg-rosti-2026}, but it remains unknown whether similar structures survive in the strongly inhomogeneous environment created by a canopy, where intense localized shear coexists with the broadband fluctuations characteristic of ET. Addressing this question experimentally is especially challenging because the relevant length scales are microscopic and the polymer stresses driving ET cannot be measured directly. Consequently, the dynamics of the viscoelastic mixing layer developing at the canopy tip remain largely unexplored.

\begin{figure}
   \includegraphics[width=\textwidth]{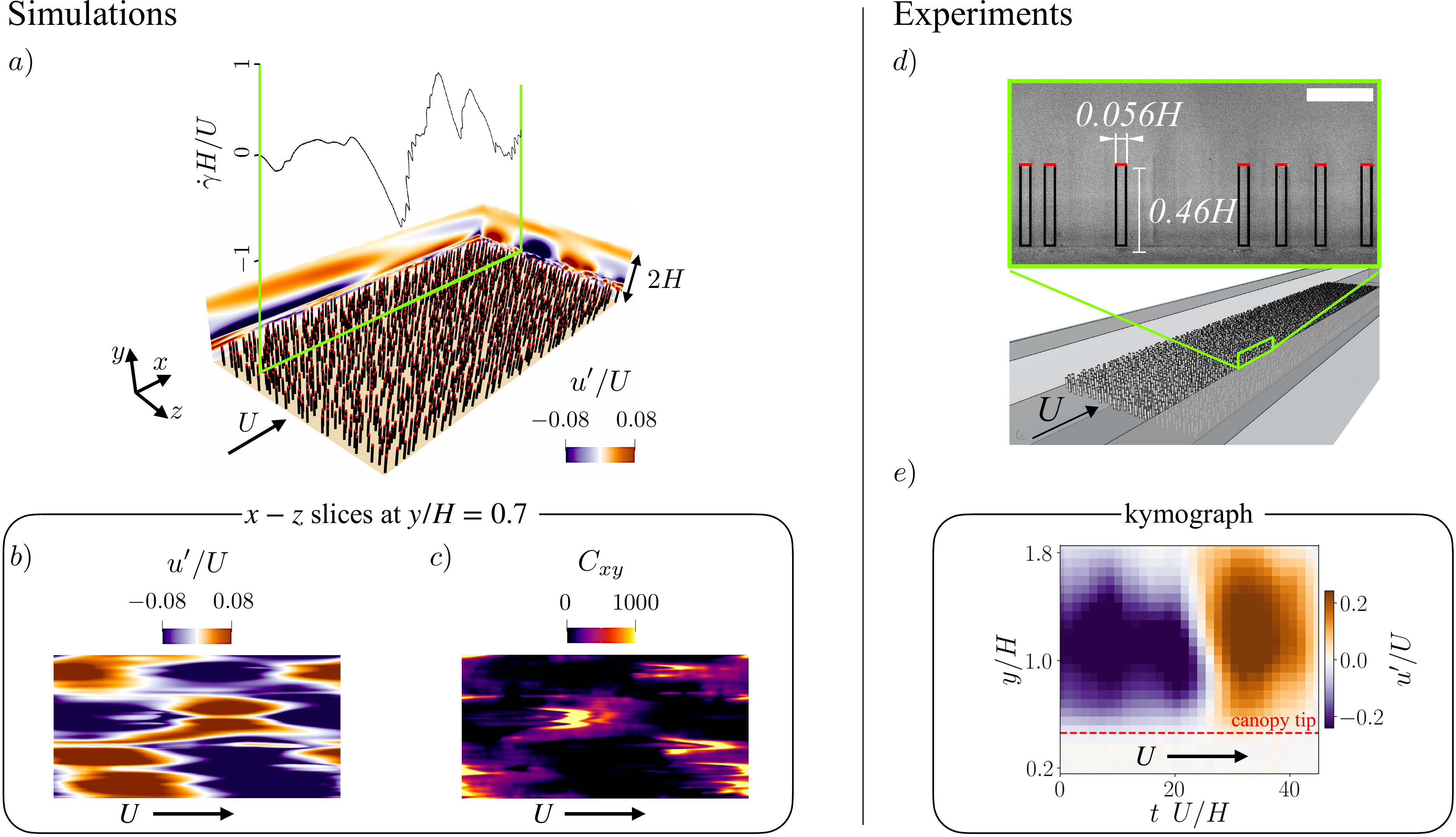}
   \caption{\textbf{Numerical simulations (left) and experiments (right) exhibit pattern formation above the canopy tip in a viscoelastic canopy flow.}
    \textbf{Panel a:} canopy array adopted in our simulations, along with snapshots of the streamwise velocity fluctuations $u^\prime$ on the sides. A streamline above the canopy tip is coloured in green, and the variation of the local shear rate $\dot{\gamma}=\partial u^\prime / \partial y$ along it is plotted above. Red dots mark pillar tips, i.e., the \textit{canopy tip}. 
    While qualitatively similar fields are observed across all of our simulations, data reported here is from the case at $Re=0.5$ with a semi-random canopy. \textbf{Panels b, c:} organised fluctuations of the streamwise velocity at $y/H=0.7$ (b), accompanied by significant polymer shearing (c). \textbf{Panel d:} computer-aided design (CAD) of the micro-channel employed in our experiments (bottom) and micrograph of the channel from a side view (top). The white scale bar corresponds to 200$\mu$m ($0.36H$, given that the channel is 1.107mm tall). \textbf{Panel e:} flow organisation emerging from the kymograph of the streamwise velocity fluctuations, sampled half-way along the canopy, at the channel centre.}  
   \label{fig:1}
\end{figure}

Here we combine fully resolved numerical simulations with microfluidic experiments to uncover the physical mechanisms governing viscoelastic flow over a rigid canopy at low Reynolds numbers. We identify a wave-mediated mechanism that organizes momentum transport in the canopy mixing layer within fully developed elastic turbulence. Specifically, we show that streamwise variations of the local shear rate (as shown in figure~\ref{fig:1}a) generate self-sustained elastic waves that persist within the chaotic flow and regulate the exchange of momentum between the free flow and the canopy. These findings reveal a previously unrecognized mechanism by which elasticity couples microscopic shear to macroscopic transport, providing a general framework for understanding how localized shear shapes the dynamics of viscoelastic fluids in structured biological and engineered environments.

\section{Results}
\subsection{Problem statement}

We consider a channel with the bottom wall covered by a dense canopy in a semi-random arrangement, protruding up to a height $h = 0.5H$, with solidity factor $\mathcal{S} = hd/\Delta S^2 \approx 0.5$ (dense canopy regime~\cite{monti-etal-2020}): $H$ denotes the channel half-height, $d$ the diameter of the pillars, and $\Delta S$ their average spacing. For comparison, we also consider a regular canopy arrangement in simulations only. See figure~\ref{fig:1} for representative views of the canopy in simulations and experiments.

Viscoelastic fluid flow within the simulated channel is characterised by the Reynolds number $Re=\rho U H / \mu_f \in \{0.5,50\}$ and the Deborah number $De=\lambda U / H=70$, along with the viscosity ratio of the polymeric solution employed $\beta=\mu_f/(\mu_f+\mu_p)=0.7$. Here, $U$ denotes the mean flow speed, $\rho$ the fluid density, $\mu_f$ the dynamic viscosity of the solvent, and $\lambda$ the polymer relaxation time. 
Elastic effects dominate over fluid inertia within the considered flow regime. 

Experiments are performed in the same dynamical range, characterised by $De/Re\gg1$, choosing {$h\approx 0.5H$, $\mathcal{S} \approx 0.35$, $Re=0.6$, $\beta \in [0.57, 0.79]$, and $De=300$ to enhance elastic effects.}

Full information on the canopy arrangements, numerical techniques and simulation domain, experimental setup and procedures are reported in the Methods section. The governing equations, quantitative matching between the computational and experimental setups, and further particulars on data analysis are provided in the Supplementary Information.

\subsection{Elastic turbulence}

\begin{figure}
   \includegraphics[width=\textwidth]{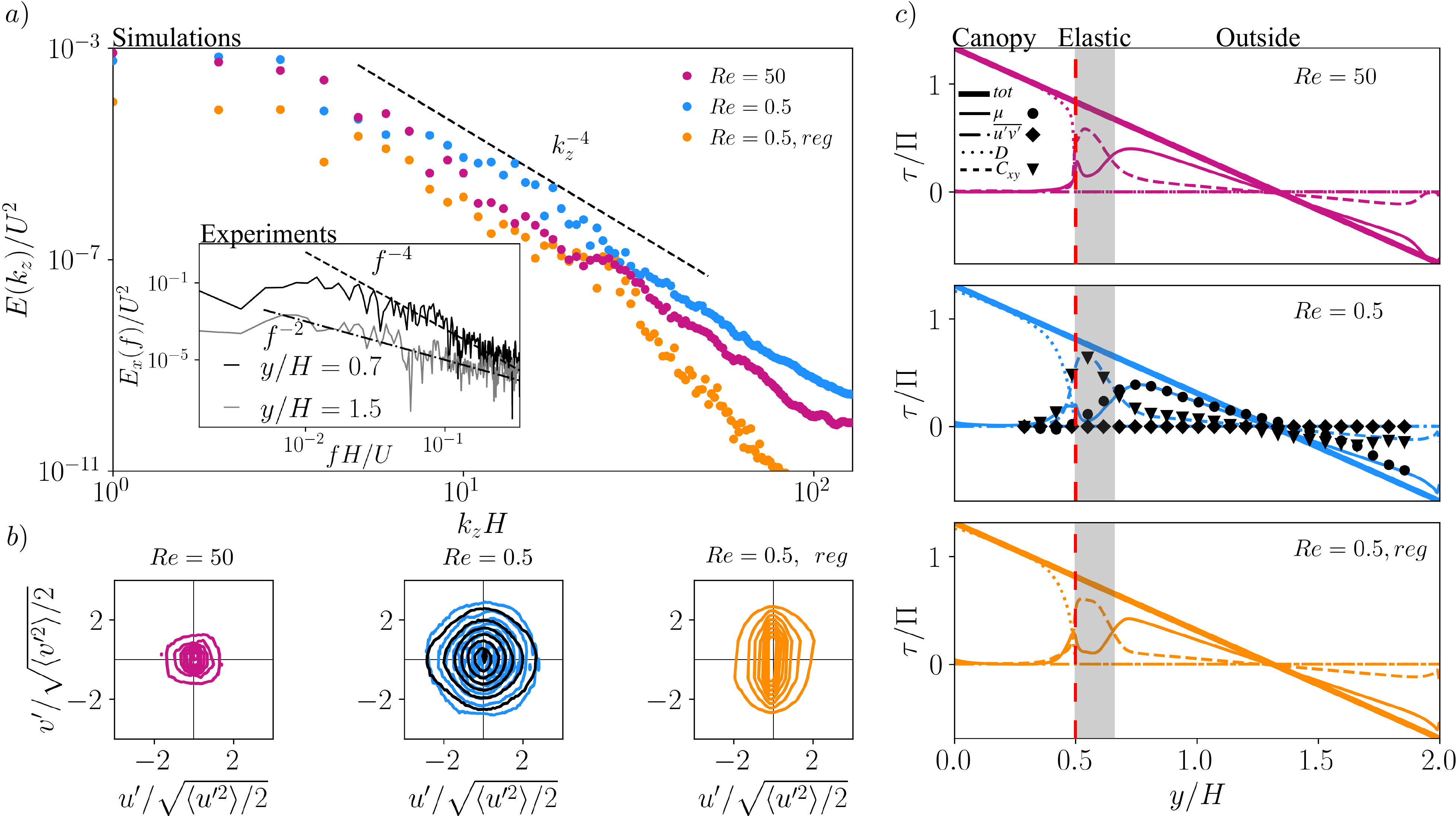}
   \caption{\textbf{Canopy elastic turbulence.} \textbf{Panel a, main axes}: spanwise spectra of the turbulent kinetic energy $E(k_z)$ above the canopy from simulations, exhibiting the $k_z^{-4}$ scaling characteristic of ET across different Reynolds numbers and canopy arrangements (here exemplified at $y/H = 0.7$, but analogous scaling behaviour is observed throughout the unobstructed channel lumen. \textit{reg} for regular canopy arrangement). \textbf{Panel a, inset}: temporal spectra of the kinetic energy in the streamwise velocity fluctuations $E_x(f)$ from experiments. Immediately above the canopy tip ($y/H = 0.7$, black line) we measure a steep $f^{-4}$ decay, while in the opposite half of the channel ($y/H = 1.5$) fluctuations amplitude is depleted and the spectrum decays as $f^{-2}$, consistent with former channel measurements \cite{lellep-linkmann-morozov-2023, foggirota-etal-2026}. \textbf{Panel b}:  joint probability density functions of streamwise and wall-normal velocity fluctuations above the canopy ($y/H = 0.7$, simulations in colours and experiments in black); their lack of correlation confirms the ET nature of the attained state \cite{foggirota-etal-2024-3,cheng-etal-2025}. \textbf{Panel c}: profiles of the contributions to the total shear stress $\tau$, time- and horizontally-averaged, normalised with the driving pressure gradient $\Pi$. Total ($tot$), viscous ($\mu$), Reynolds ($\overline{u^\prime v^\prime}$), canopy drag ($D$), and polymeric ($C_{xy}$); curves from simulations and symbols from experiments (experiments are actually performed at $Re=0.6$, and compared to the simulation with the closest matching $Re$). Polymeric shear from experiments is found by closure, using drag measurements from simulations; details in the Supplementary Information. Grey shading immediately above the canopy tip (red dashed line) identifies the elastically active layer, separating the canopy from the outside. Reynolds shear is essentially null throughout, consistent with ET, while remaining contributions expose the dominant role of polymeric shear within the elastically active layer and underpin the identification of the three flow regions.}
   \label{fig:2}
\end{figure}

Sustained velocity fluctuations populate the flow over the canopy and permeate all length scales, as confirmed by the turbulent kinetic energy spectra in the main axes of figure~\ref{fig:2}a. 
Their steep $k^{-4}$ scaling in the wavenumbers $k$, that is the hallmark of ET \cite{foggirota-etal-2026}, emerges consistently across the different Reynolds numbers and canopy arrangements considered in our simulations (results in figure~\ref{fig:2} are from the semi-random canopy, unless labelled with \textit{reg} to denote the regular one). Experimental measurements of the streamwise velocity spectra in the frequencies $f$ (figure~\ref{fig:2}a, inset) match the $f^{-2}$ scaling reported for channels \cite{lellep-linkmann-morozov-2023, foggirota-etal-2026} far off from the canopy (grey curve), but grow in magnitude and attain a steeper $f^{-4}$ scaling in proximity of its tip (black curve).
The local matching of the temporal and spatial scaling exponents immediately above the canopy highlights peculiar ongoing dynamics there, and is fully understood in the following section.

The ET nature of the attained chaotic state is further confirmed by the absence of correlation between streamwise and wall-normal velocity fluctuations \cite{foggirota-etal-2024-3,cheng-etal-2025}, apparent in the joint probability density functions (J-PDFs) shown in figure~\ref{fig:2}b (simulations in colours, experiments in black).

After establishing that the flow satisfies the canonical signatures of ET, we identify previously unrecognised polymeric activity immediately above the canopy tip.
In fact, three markedly different flow regions are revealed by the different contributions to the mean shear stress balance across the channel, reported in figure~\ref{fig:2}c for all simulated cases (continuous lines).
Flow inside the canopy is dominated by drag, consistent with canopy flow literature at both high and low Reynolds numbers \cite{poggi-etal-2004, lopez-haward-shen-2025}. 
The outer flow is instead governed by viscous shear and bears strong similarities with ET in channel flows \cite{foggirota-etal-2024-3}, with quasi-linear stress profiles exhibiting only minor deviations near the upper wall, attributable to the polymeric diffusive instability \cite{beneitez-page-kerswell-2023}.
Most remarkably, our analysis reveals the formation of an \textit{elastically active layer} immediately above the canopy tip, in which polymeric shear dominates all other stress contributions.
Additional black symbols overlaid to the $Re=0.5$ case represent experimental measurements and appear to fully support this picture. 
The layer of localised polymeric activity, bridging the mixing-layer and transitional-flow regions previously isolated in viscoelastic canopy flows \cite{lopez-haward-shen-2025}, is the focus of the next section.

\subsection{Elastic waves}

\begin{figure}
   \includegraphics[width=\textwidth]{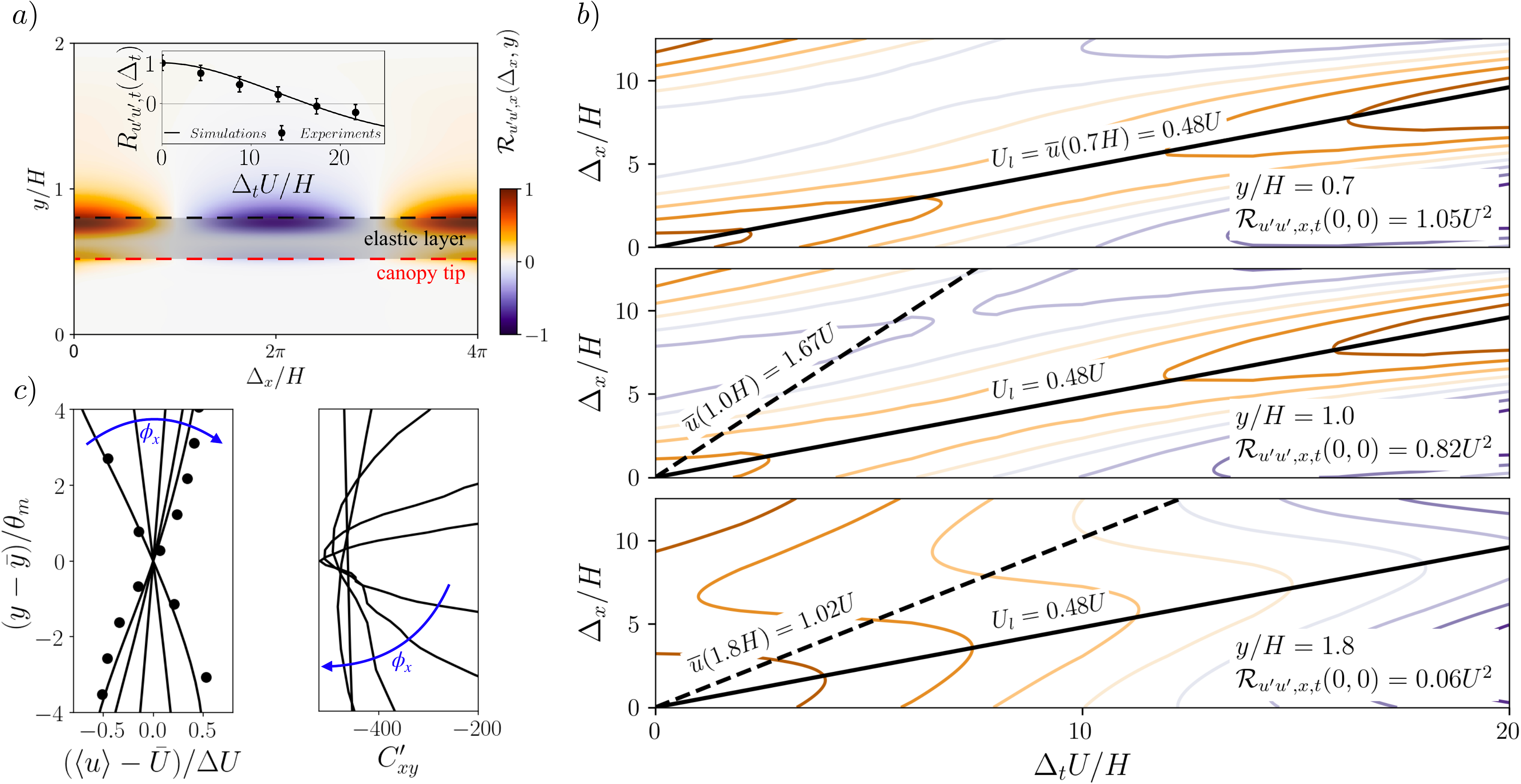}
   \caption{\textbf{Wave propagation over the canopy tip.}
   \textbf{Panel a, main axes}: streamwise autocorrelation $\mathcal{R}_{u^\prime u^\prime,x}$ of the streamwise velocity fluctuations $u^\prime$ as a function of the wall-normal coordinate $y$ and the streamwise separation $\Delta_x$, from the simulation with a semi-random canopy at $Re=0.5$ (shown throughout this figure); the signal is confined in proximity of the elastically active layer. The canopy tip is denoted with a red dashed line, while the black dashed line highlights the location from which the inset data is extracted. \textbf{Panel a, inset}: the simulated autocorrelation at $y/H=0.7$ (continuous line), properly converted into a function of the time separation $\mathcal{R}_{u^\prime u^\prime,t}$ (details in the Supplementary Information), is compared to the experimentally measured autocorrelation (dots), sampled at the canopy centre and averaged within the elastically active layer. The good agreement in correlation shape confirms that simulations and experiments capture analogous spatial and temporal patterns. Error bars denote one standard deviation.
\textbf{Panel b}: space-time autocorrelation $\mathcal{R}_{u^\prime u^\prime,x,t}$ of the streamwise velocity fluctuations $u^\prime$ sampled at $y/H\in\{ 0.7, 1, 1.8 \}$. Contour lines from orange to purple denote ten evenly-spaced levels between the correlation maximum and minimum, while the un-normalised maximum value $\mathcal{R}_{u^\prime u^\prime,x,t}(0,0)$ is shown in the legend. Correlation peaks are seen to follow a continuous line with slope $U_l$ corresponding to the mean advection of the elastically active layer, regardless of the $y$ position. The local advection velocity $\overline{u}(y)$ is also shown for comparison as a dashed line at $y/H\in\{1, 1.8 \}$, while $y/H=0.7$ lies within the layer and thus $U_l=\overline{u}$. \textbf{Panel c}: reversal of the phase-averaged velocity profile (left) and accompanying trend of the polymeric shear (right) in generalised mixing layer coordinates \cite{ghisalberti-nepf-2002}: $\theta_m$ is the momentum-thickness of the mixing layer, $\Delta U$ the velocity variation across it, and barred quantities are averaged throughout. Continuous lines denote simulation data and span a phase range $\phi_x\in(0,0.06\pi)$, while circular markers are experimental measurements at a time shift of $5 H/U$; both are acquired at the canopy centre. Error bars for experimental data are smaller than the marker size.}
   \label{fig:3}
\end{figure}

The horizontal slices of the flow in figure~\ref{fig:1}b,c reveal the formation of a pattern in the fluctuations of the streamwise velocity component and in the viscoelastic shear, consistently observed across all investigated cases since dominated by elasticity ($De/Re\gg1$). Results from our simulation with a semi-random canopy at $Re=0.5$ are thus representatively shown in the following.
The spatial structure of the pattern is described by the streamwise autocorrelation of the velocity fluctuations, reported in figure~\ref{fig:3}a: a strong signal (i.e., spatial organisation of the flow) emerges exclusively in the immediate proximity of the elastically active layer.
The correlation peaks span a thicker layer compared to that emerging from the polymeric shear in figure~\ref{fig:2}c, suggesting that the localised elastic dynamics affect the fluid on a broader vertical extent, not limited to the region where the polymeric shear stress dominates.
Good agreement of the correlation shape is found between simulations and experiments (in the inset), confirming that both capture fluctuations with comparable correlation structure.

We then observe the space-time autocorrelation at selected stations above the canopy, in figure~\ref{fig:3}b. The top panel, reporting measurements inside the layer, shows how the advection velocity of the pattern ($U_l$) corresponds to the average flow speed there, $\overline{u} (y)$. Remarkably, fluctuations are observed to propagate with speed $U_l$ also at other vertical positions (reported below), where the local value of $\overline{u} (y)$ is instead different. This behaviour, similar to that induced by Kelvin-Helmholtz rollers in high-Re mixing layers \cite{finnigan-2000}, highlights how the dynamics generated inside the elastically active layer effectively influences the whole flow above. Nevertheless, the coherence of the pattern itself is depleted moving away from the layer, as confirmed by the decreasing values of the un-normalised autocorrelation.

We interpret pattern formation as a manifestation of the elastic waves identified by Foggi Rota \textit{et al.} \cite{foggirota-garg-rosti-2026} for the temporal evolution of a one-dimensional mixing layer.
There, it was shown that a freely decaying mixing layer of a viscoelastic fluid spontaneously undergoes periodic flow reversals -- i.e., \textit{yo-yos}, as opposed to the monotonic decay of the Newtonian case.
The yo-yoing of the mean flow is driven by the flow shear rotating the elastic polymers, which in turn inject energy back into the fluid.
These peculiar dynamics have been mathematically modelled and treated analytically, finding solutions in the form of standing waves, with fixed structure along the gradient direction and pulsating in time.
Such waves lead the fluctuating velocity to alternate between positive and negative values through the coupled, out-of-phase evolution of the polymeric shear. In the present setup, the problem has more spatial dimensions, and the role of temporal anisotropy is taken by the spatial inhomogeneity of the canopy along the streamwise direction, while structure along the gradient direction is retained. To confirm this conjecture, we construct a phase-averaged field $\langle u \rangle (\phi_x,y,z)$, where $\phi_x = 2\pi (x - U_l t)$ is the streamwise phase and $t$ denotes time. At multiple locations within the elastically active layer, upon rescaling $\langle u \rangle$ in generalised mixing layer coordinates \cite{ghisalberti-nepf-2002} (details in the Supplementary Information), we find velocity reversals in strong analogy with the theoretical arguments \cite{foggirota-garg-rosti-2026}, as exemplified in figure~\ref{fig:3}c. The mixing layer is observed to yo-yo in a phase range comparable to the shear modulation (here imposed by the mean spacing between the pillars), while the polymeric shear is out of phase from the reversals, thus sustaining the cycle. The reversing velocity profile appears comparable between simulation data (continuous lines) and experimental measurements (circles).

Finally, we are now optimally placed to better understand the trends of the temporal spectra reported in the inset of figure~\ref{fig:2}a. 
In general, no linear transformation can be found to relate spatial and temporal measurements in pure ET \cite{foggirota-etal-2026} (i.e., as a rule, \textit{Taylor's frozen turbulence hypothesis} \cite{taylor-1938} does not hold).
Yet, within the elastically active layer, pattern propagation at the mean advection locally provides the unique velocity scale needed for the matching; the signature spatial scaling of ET is thus recovered there also in the frequencies.
Outside the layer, instead, local advection diverges from the pattern propagation velocity, leading to the observed multi-scaling.

\section{Discussion}

Our results reveal that the statistical disorder of elastic turbulence and the coherent motion of elastic waves
can coexist in viscoelastic shear flow,
sustained by the same underlying elasticity--shear interaction.
In viscoelastic canopy flow, the strong spatial heterogeneity introduced by the canopy elements generates an elastically active layer immediately above the canopy tip that both sustains the chaotic dynamics characteristic of elastic turbulence and supports the propagation of elastic waves. 

The observed wave propagation extends recent theoretical developments on elasticity--shear coupling \cite{foggirota-garg-rosti-2026} from temporally varying to spatially heterogeneous flows. In the idealised theoretical setting, temporal variations of the background shear provide the conditions required for wave propagation. Here, the canopy naturally generates the corresponding spatial variability, establishing the same physical mechanism without externally imposed forcing. The robustness of the phenomenon across different Reynolds numbers and both ordered and random canopy arrangements further suggests that it is not a consequence of a particular geometry, but rather a generic feature of viscoelastic flows with sufficiently heterogeneous shear.

Within the elastically active layer, wave propagation locally establishes a linear correspondence between space and time, recovering the mapping underlying Taylor's frozen-turbulence hypothesis \cite{taylor-1938}. This correspondence reconciles the temporal and spatial scalings of the turbulent spectra despite the general breakdown of Taylor's hypothesis in elastic turbulence \cite{foggirota-etal-2026}. Importantly, this local space--time equivalence is confined to the wave-supporting layer, even though the influence of the waves extends throughout much of the flow above the canopy, highlighting the non-local impact of elasticity on the surrounding turbulent dynamics.

More broadly, these findings suggest that elastic wave propagation may be a generic feature of viscoelastic flows with strong spatial heterogeneity, rather than a peculiarity of idealised configurations. Localised regions of intense shear are ubiquitous in biological transport systems, porous media, rough surfaces, vegetated environments, and microfluidic devices, raising the possibility that coherent elastic waves coexist with elastic turbulence in a much wider class of flows than what previously recognised. This perspective motivates the development of a unified theoretical framework describing elasticity in spatially heterogeneous flows.
Beyond their fundamental significance, our results suggest that flow geometry itself can be exploited to trigger the onset of chaotic dynamics and coherent elastic waves, opening new opportunities for manipulating mixing, transport, and for characterising viscoelastic fluids in biological and engineered systems.

\section*{Methods}

\subsection*{Simulations}

Simulating viscoelastic fluid flow in the elastic turbulence regime presents substantial numerical challenges. The governing equations require high-order spatial and temporal discretisation to faithfully capture the multi-scale dynamics of the polymer-laden flow, and the computational cost of adequate resolution makes such simulations a significant undertaking \cite{dubief-terrapon-hof-2023}. All simulations reported in this work were carried out with the solver \textit{Fujin} \cite{rosti-2026} (\url{https://www.oist.jp/research/research-units/cffu/fujin}), developed and maintained by our unit at OIST, and extensively validated for canopy and viscoelastic flow problems \cite{foggirota-etal-2024-2, foggirota-etal-2024-3, foggirota-etal-2026}. The incompressible Navier-Stokes equations are discretised on a staggered uniform Cartesian grid using a second-order central finite-difference scheme in space, and integrated in time with a second-order Adams--Bashforth scheme coupled to a fractional step method \cite{kim-moin-1985}. At each time step, the divergence-free pressure field is obtained via a spectral Poisson solver, and parallelisation is achieved through the 2DECOMP library with MPI. 
We describe the internal fluid microstructure through the Oldroyd-B model \cite{oldroyd-1950}, and solve the conformation tensor equation in its logarithmic formulation \cite{fattal-kupferman-2005, devita-etal-2018} to circumvent the high-Deborah numerical instability, with the upper-convected derivative treated with a weighted essentially non-oscillatory (WENO) scheme \cite{sugiyama-etal-2011}. This combination avoids the need for an explicit stress-diffusion term, and consequently no boundary condition on the conformation tensor is required at the walls \cite{beneitez-etal-2024} or at the canopy pillars. No-slip and no-penetration conditions are applied to the fluid at the upper and lower walls, and periodicity is enforced in the streamwise and spanwise directions. The fluid and the canopy pillars are coupled at their interface through a no-slip and no-penetration boundary condition, guaranteed applying the force distribution computed with a Lagrangian immersed boundary method \cite{peskin-2002, huang-etal-2007, banaei-rosti-brandt-2020} along 160 points per pillar, homogeneously distributed in line.

The canopy consists of 1352 pillars of diameter $d \approx 0.06H$, arranged in either a semi-random or regular configuration. In the semi-random case, the channel floor is partitioned into a $25 \times 52$ grid of rectangular tiles of area $\Delta S^2$, and one pillar is placed within each tile at a position drawn from a uniform distribution; this tiling serves solely to control the canopy parameters and bears no relation to the computational grid. In the regular case, each pillar is placed at the centre of its tile. All simulations use a computational domain of dimensions $4\pi H \times 2H \times 2\pi H$ in the streamwise, wall-normal, and spanwise directions respectively, discretised on a uniform grid of $512 \times 512 \times 256$ points. Although no prior benchmarks exist for viscoelastic canopy flow simulations specifically, this domain size and resolution have been shown to adequately capture and resolve the relevant flow structures in elastic turbulence channel flow, both in our own previous work \cite{foggirota-etal-2024-3, foggirota-etal-2026} and in other studies \cite{deangelis-casciola-piva-2002, lellep-linkmann-morozov-2024}. Confinement effects and their circumvention through experimental measurements are discussed in the Supplementary Information.

A total of three simulations were performed: two at $Re = 0.5$ with the semi-random and regular canopy arrangements, and one at $Re = 50$ with the semi-random arrangement.
Starting from a fully developed turbulent state at high Reynolds, we left transients elapse and collected the statistics reported in the main text at stationary state, for a time span of $\sim\lambda$, i.e. $\sim 70 H/U$. Each simulation was run on 8192 Fujitsu A64FX cores of the Fugaku supercomputer at the RIKEN Center for Computational Science in Kobe, Japan, with a wall-clock execution time of approximately six months per simulation.

\subsection*{Experiments}

The reported experimental results are obtained in a straight rectangular micro-channel fabricated in fused silica by the selective laser etching (SLE) method using a LightFab 3D printer (LightFab, GmbH). The inner dimensions of the channel are $867.2 H \times 2H \times 7.6H$ in the streamwise, wall-normal, and spanwise directions, respectively, with $H=0.554$mm. The canopy consists of 3591 pillars, each with a diameter $d\approx0.056H \pm 0.003H$, a height $h \approx 0.46H \pm 0.007H$ (approximately 1/2 of the channel half-height), and an average spacing of $\Delta S = 0.27H$, resulting in a solidity factor $\mathcal{S}=0.35 \pm 0.03$. The channel floor is partitioned into a $27 \times 133$ grid of square tiles of area $\Delta S^2$, and each pillar is positioned within each tile at a random position drawn from a uniform distribution, redrawing the position if any pillar happens to overlap any of its neighbours. The canopy expands for $36H$ in the streamwise direction, starting at $21.7H$ downstream of the channel inlet, and covers the entire channel span.

The test fluid is a polymer solution of polyacrylamide (5 MDa, Sigma-Aldrich) at a concentration of 400 ppm in a glycerol-water mixture (e-Nacalai and Milli-Q) with an $89.3$ wt$\%$ glycerol content. A 5085 ppm solution of polyacrylamide was prepared by dissolving the polymer powder in water using a roller mixer until the measured relaxation time became consistent across samples taken from the top, middle, and bottom of the container. The aqueous solution was then mixed overnight with the required mass of glycerol using a roller mixer. The refractive index of the solution is $n_\mathrm{ref}=1.4581$, measured with an Abbemat MW refractometer (589 nm, Anton Paar), which closely matches the refractive index of the fused silica device. The viscosity is measured in a stress-controlled shear rheometer (MCR 502, Anton Paar GmbH) at 25 \textdegree C using a cone-plate geometry (50 mm, 1\textdegree angle). The fluid is very weakly shear-thinning with an average viscosity of 0.2 Pas (the flow curve is shown in the Supplementary Information). The relaxation time $\lambda = 0.78 \pm 0.01$ s is measured using a capillary breakup extensional rheometer (Haake CaBER, Thermo Fisher Scientific), at 25\textdegree C with 6 mm plates, initial height of 1 mm, and a Hencky strain of 1.8, and a separation speed of 25 mm/s.

The velocity is measured at the middle of the canopy using a micro-Tomographic Particle Image Velocimetry ($\mu$-TPIV) system (LaVision FlowMaster) comprising a stereomicroscope (SteREO V20, Zeiss AG), two high-speed cameras (1280 × 800 pixels, Phantom VEO 410), and a pulsed Nd:YLF laser (527 nm wavelength, Photonics Industries). The solution was seeded with fluorescent tracer particles (3.2 $\mu$m, polystyrene, Fluoromax red, Thermo Scientific Inc). Their three-dimensional location is determined after applying a reconstruction algorithm part of the DaVis software from LaVision, starting from a 2D field in the streamwise/wall-normal direction (a more detailed description can be found in our former work \cite{carlson-shen-haward-2021}). The flow is controlled using a single syringe pump (Nemesys S, Cetoni GmbH) loaded with glass syringes (1000 Series Gastight Syringes, Hamilton Company). The channel outlet is connected to a long hose, open at its end to the atmosphere. This way, instabilities at the channel outlet are prevented from traveling upstream and affecting the canopy flow.

\renewcommand{\bibsection}{{ \section*{References}}}
\bibliographystyle{naturemag}
\bibliography{Wallturb}

@article{banaei-rosti-brandt-2020,
  title = {Numerical Study of Filament Suspensions at Finite Inertia},
  author = {Banaei, A. A. and Rosti, M. E. and Brandt, L.},
  year = {2020},
  journal = {J. Fluid Mech.},
  volume = {882},
  pages = {A5},
  issn = {0022-1120, 1469-7645},
  doi = {10.1017/jfm.2019.794},
  urldate = {2023-04-13},
  langid = {english}
}

@article{beneitez-etal-2024,
  title = {Multistability of Elasto-Inertial Two-Dimensional Channel Flow},
  author = {Beneitez, M. and Page, J. and Dubief, Y. and Kerswell, R. R.},
  year = {2024},
  month = feb,
  journal = {J. Fluid Mech.},
  volume = {981},
  pages = {A30},
  issn = {0022-1120, 1469-7645},
  doi = {10.1017/jfm.2024.50},
  urldate = {2024-03-25},
  langid = {english}
}

@article{beneitez-page-kerswell-2023,
  title = {Polymer Diffusive Instability Leading to Elastic Turbulence in Plane {{Couette}} Flow},
  author = {Beneitez, M. and Page, J. and Kerswell, R. R.},
  year = {2023},
  month = oct,
  journal = {Phys. Rev. Fluids},
  volume = {8},
  number = {10},
  pages = {L101901},
  publisher = {American Physical Society},
  doi = {10.1103/PhysRevFluids.8.L101901},
  urldate = {2023-11-16}
}

@book{bird-1987,
  title = {Dynamics of {{Polymeric Liquids}}, {{Volume}} 1: {{Fluid Mechanics}}},
  shorttitle = {Dynamics of {{Polymeric Liquids}}, {{Volume}} 1},
  author = {Bird, R. B.},
  year = {1987},
  publisher = {Wiley},
  isbn = {978-0-471-80245-7},
  langid = {english}
}

@article{carlson-shen-haward-2021,
  title = {Microtomographic Particle Image Velocimetry Measurements of Viscoelastic Instabilities in a Three-Dimensional Microcontraction},
  author = {Carlson, D. W. and Shen, A. Q. and Haward, S. J.},
  year = {2021},
  month = sep,
  journal = {J. Fluid Mech.},
  volume = {923},
  pages = {R6},
  issn = {0022-1120, 1469-7645},
  doi = {10.1017/jfm.2021.620},
  urldate = {2026-07-10},
  langid = {english}
}

@article{cheng-etal-2025,
  title = {Anomalous {{Reynolds}} Stress and Dynamic Mechanisms in Two-Dimensional Elasto-Inertial Turbulence of Viscoelastic Channel Flow},
  author = {Cheng, H. and Zhang, H. and Zhang, W. and Wang, S. and Li, X. and Li, Y. and Li, F. C.},
  year = {2025},
  month = sep,
  journal = {J. Fluid Mech.},
  volume = {1018},
  pages = {A33},
  issn = {0022-1120, 1469-7645},
  doi = {10.1017/jfm.2025.10538},
  urldate = {2026-06-17},
  langid = {english}
}

@article{deangelis-casciola-piva-2002,
  title = {{{DNS}} of Wall Turbulence: Dilute Polymers and Self-Sustaining Mechanisms},
  shorttitle = {{{DNS}} of Wall Turbulence},
  author = {De Angelis, E. and Casciola, C. M. and Piva, R.},
  year = {2002},
  journal = {Comput. Fluids},
  volume = {31},
  number = {4},
  pages = {495--507},
  issn = {0045-7930},
  urldate = {2021-10-31},
  langid = {english}
}

@article{deblois-haward-shen-2023,
  title = {Canopy Elastic Turbulence: {{Spontaneous}} Formation of Waves in Beds of Slender Microposts},
  shorttitle = {Canopy Elastic Turbulence},
  author = {{de Blois}, C. and Haward, S. J. and Shen, A. Q.},
  year = {2023},
  month = feb,
  journal = {Phys. Rev. Fluids},
  volume = {8},
  number = {2},
  pages = {023301},
  publisher = {American Physical Society},
  doi = {10.1103/PhysRevFluids.8.023301},
  urldate = {2023-10-25}
}

@article{devita-etal-2018,
  title = {Elastoviscoplastic Flows in Porous Media},
  author = {De Vita, F. and Rosti, M. E. and Izbassarov, D. and Duffo, L. and Tammisola, O. and Hormozi, S. and Brandt, L.},
  year = {2018},
  month = aug,
  journal = {J. Nonnewton. Fluid. Mech.},
  volume = {258},
  pages = {10--21},
  issn = {0377-0257},
  doi = {10.1016/j.jnnfm.2018.04.006},
  urldate = {2023-06-19},
  langid = {english}
}

@article{dubief-terrapon-hof-2023,
  title = {Elasto-{{Inertial Turbulence}}},
  author = {Dubief, Y. and Terrapon, V. E. and Hof, B.},
  year = {2023},
  journal = {Annu. Rev. Fluid Mech.},
  volume = {55},
  number = {1},
  pages = {675--705},
  doi = {10.1146/annurev-fluid-032822-025933},
  urldate = {2023-06-19}
}

@article{fattal-kupferman-2005,
  title = {Time-Dependent Simulation of Viscoelastic Flows at High {{Weissenberg}} Number Using the Log-Conformation Representation},
  author = {Fattal, R. and Kupferman, R.},
  year = {2005},
  month = feb,
  journal = {J. Nonnewton. Fluid. Mech.},
  volume = {126},
  number = {1},
  pages = {23--37},
  issn = {0377-0257},
  doi = {10.1016/j.jnnfm.2004.12.003},
  urldate = {2023-06-19},
  langid = {english}
}

@article{finnigan-2000,
  title = {Turbulence in {{Plant Canopies}}},
  author = {Finnigan, J.},
  year = {2000},
  journal = {Annu. Rev. Fluid Mech.},
  volume = {32},
  number = {1},
  pages = {519--571},
  doi = {10.1146/annurev.fluid.32.1.519},
  urldate = {2023-02-01}
}

@article{foggirota-etal-2024-2,
  title = {Dynamics and Fluid--Structure Interaction in Turbulent Flows within and above Flexible Canopies},
  author = {Foggi Rota, G. and Monti, A. and Olivieri, S. and Rosti, M. E.},
  year = {2024},
  journal = {J. Fluid Mech.},
  volume = {989},
  pages = {A11},
  issn = {0022-1120, 1469-7645},
  doi = {10.1017/jfm.2024.481},
  urldate = {2024-07-29},
  langid = {english}
}

@article{foggirota-etal-2024-3,
  title = {Unified View of Elastic and Elasto-Inertial Turbulence in Channel Flows at Low and Moderate {{Reynolds}} Numbers},
  author = {Foggi Rota, G. and Amor, C. and Le Clainche, S. and Rosti, M. E.},
  year = {2024},
  journal = {Phys. Rev. Fluids},
  volume = {9},
  number = {12},
  pages = {L122602},
  publisher = {American Physical Society},
  doi = {10.1103/PhysRevFluids.9.L122602},
  urldate = {2025-01-03}
}

@article{foggirota-etal-2026,
  title = {The Broken Link between Space and Time in Elastic Turbulence},
  author = {Foggi Rota, G. and Singh, R. K. and Chiarini, A. and Amor, C. and Soligo, G. and Mitra, D. and Rosti, M. E.},
  year = {2026},
  month = mar,
  journal = {Int. J. Multiph. Flow},
  volume = {197},
  pages = {105630},
  issn = {0301-9322},
  doi = {10.1016/j.ijmultiphaseflow.2026.105630},
  urldate = {2026-02-20}
}

@article{foggirota-garg-rosti-2026,
  title = {Waves Dictate the Yo-Yoing Decay of a Viscoelastic Mixing Layer},
  author = {Foggi Rota, G. and Garg, P. and Tang, J. and Rosti, M. E.},
  year = {2026},
  month = jul,
  journal = {Commun. Phys.},
  volume = {(in press)},
  publisher = {Nature Publishing Group},
  issn = {2399-3650},
  doi = {10.1038/s42005-026-02748-8},
  urldate = {2026-07-14},
  copyright = {2026 The Author(s)},
  langid = {english}
}

@article{ghisalberti-nepf-2002,
  title = {Mixing Layers and Coherent Structures in Vegetated Aquatic Flows},
  author = {Ghisalberti, M. and Nepf, H. M.},
  year = {2002},
  journal = {J. Geophys. Res. Oceans},
  volume = {107},
  number = {C2},
  pages = {3-1-3-11},
  issn = {2156-2202},
  doi = {10.1029/2001JC000871},
  urldate = {2023-05-02},
  langid = {english}
}

@article{groisman-steinberg-2000,
  title = {Elastic Turbulence in a Polymer Solution Flow},
  author = {Groisman, A. and Steinberg, V.},
  year = {2000},
  month = may,
  journal = {Nature},
  volume = {405},
  number = {6782},
  pages = {53--55},
  publisher = {Nature Publishing Group},
  issn = {1476-4687},
  doi = {10.1038/35011019},
  urldate = {2023-06-19},
  copyright = {2000 Macmillan Magazines Ltd.},
  langid = {english}
}

@article{groisman-steinberg-2001-2,
  title = {Efficient Mixing at Low {{Reynolds}} Numbers Using Polymer Additives},
  author = {Groisman, A. and Steinberg, V.},
  year = {2001},
  month = apr,
  journal = {Nature},
  volume = {410},
  number = {6831},
  pages = {905--908},
  publisher = {Nature Publishing Group},
  issn = {1476-4687},
  doi = {10.1038/35073524},
  urldate = {2023-06-19},
  copyright = {2001 Macmillan Magazines Ltd.},
  langid = {english}
}

@article{huang-etal-2007,
  title = {Simulation of Flexible Filaments in a Uniform Flow by the Immersed Boundary Method},
  author = {Huang, W. X. and Shin, S. J. and Sung, H. J.},
  year = {2007},
  journal = {J. Comput. Phys.},
  volume = {226},
  number = {2},
  pages = {2206--2228},
  issn = {0021-9991},
  doi = {10.1016/j.jcp.2007.07.002},
  urldate = {2022-12-01},
  langid = {english}
}

@article{kim-moin-1985,
  title = {Application of a Fractional-Step Method to Incompressible {{Navier-Stokes}} Equations},
  author = {Kim, J. and Moin, P.},
  year = {1985},
  journal = {J. Comput. Phys.},
  volume = {59},
  number = {2},
  pages = {308--323},
  publisher = {Elsevier},
  doi = {10.1016/0021-9991(85)90148-2}
}

@article{lellep-linkmann-morozov-2023,
  title = {Linear Stability Analysis of Purely Elastic Travelling-Wave Solutions in Pressure-Driven Channel Flows},
  author = {Lellep, M. and Linkmann, M. and Morozov, A.},
  year = {2023},
  month = mar,
  journal = {J. Fluid Mech.},
  volume = {959},
  pages = {R1},
  publisher = {Cambridge University Press},
  issn = {0022-1120, 1469-7645},
  doi = {10.1017/jfm.2023.100},
  urldate = {2023-06-19},
  langid = {english}
}

@article{lellep-linkmann-morozov-2024,
  title = {Purely Elastic Turbulence in Pressure-Driven Channel Flows},
  author = {Lellep, M. and Linkmann, M. and Morozov, A.},
  year = {2024},
  month = feb,
  journal = {PNAS},
  volume = {121},
  number = {9},
  pages = {e2318851121},
  publisher = {Proceedings of the National Academy of Sciences},
  doi = {10.1073/pnas.2318851121},
  urldate = {2024-03-05}
}

@article{loiseau-etal-2020,
  title = {Active Mucus--Cilia Hydrodynamic Coupling Drives Self-Organization of Human Bronchial Epithelium},
  author = {Loiseau, E. and Gsell, S. and Nommick, A. and Jomard, C. and Gras, D. and Chanez, Pascal and D'Ortona, Umberto and Kodjabachian, Laurent and Favier, Julien and Viallat, Annie},
  year = {2020},
  journal = {Nat. Phys.},
  volume = {16},
  number = {11},
  pages = {1158--1164},
  publisher = {Nature Publishing Group},
  issn = {1745-2481},
  doi = {10.1038/s41567-020-0980-z},
  urldate = {2023-02-02},
  copyright = {2020 The Author(s), under exclusive licence to Springer Nature Limited},
  langid = {english}
}

@article{lopez-haward-shen-2025,
  title = {Canopy Elastic Turbulence: {{Insights}} and Analogies to Canopy Inertial Turbulence},
  shorttitle = {Canopy Elastic Turbulence},
  author = {{Lopez de la Cruz}, R. and Haward, S. J. and Shen, A. Q.},
  year = {2025},
  month = jan,
  journal = {PNAS Nexus},
  volume = {4},
  number = {1},
  pages = {pgae571},
  issn = {2752-6542},
  doi = {10.1093/pnasnexus/pgae571},
  urldate = {2026-02-19}
}

@article{monti-etal-2020,
  title = {On the Genesis of Different Regimes in Canopy Flows: A Numerical Investigation},
  shorttitle = {On the Genesis of Different Regimes in Canopy Flows},
  author = {Monti, A. and Omidyeganeh, M. and Eckhardt, B. and Pinelli, A.},
  year = {2020},
  journal = {J. Fluid Mech.},
  volume = {891},
  pages = {A9},
  issn = {0022-1120, 1469-7645},
  doi = {10.1017/jfm.2020.155},
  urldate = {2022-03-25},
  langid = {english}
}

@article{morozov-2022,
  title = {Coherent {{Structures}} in {{Plane Channel Flow}} of {{Dilute Polymer Solutions}} with {{Vanishing Inertia}}},
  author = {Morozov, A.},
  year = {2022},
  month = jun,
  journal = {Phys. Rev. Lett.},
  volume = {129},
  number = {1},
  pages = {017801},
  publisher = {American Physical Society},
  doi = {10.1103/PhysRevLett.129.017801},
  urldate = {2023-05-30}
}

@article{nepf-2012-1,
  title = {Flow and {{Transport}} in {{Regions}} with {{Aquatic Vegetation}}},
  author = {Nepf, H. M.},
  year = {2012},
  journal = {Annu. Rev. Fluid Mech.},
  volume = {44},
  number = {1},
  pages = {123--142},
  issn = {0066-4189, 1545-4479},
  doi = {10.1146/annurev-fluid-120710-101048},
  urldate = {2022-03-25},
  langid = {english}
}

@article{oldroyd-1950,
  title = {On the Formulation of Rheological Equations of State},
  author = {Oldroyd, J. G.},
  year = {1950},
  journal = {Proc. R. Soc. Lond. A},
  volume = {200},
  number = {1063},
  pages = {523--541},
  publisher = {Royal Society},
  doi = {10.1098/rspa.1950.0035},
  urldate = {2023-10-31}
}

@article{page-dubief-kerswell-2020,
  title = {Exact {{Traveling Wave Solutions}} in {{Viscoelastic Channel Flow}}},
  author = {Page, J. and Dubief, Y. and Kerswell, R. R.},
  year = {2020},
  month = oct,
  journal = {Phys. Rev. Lett.},
  volume = {125},
  number = {15},
  pages = {154501},
  publisher = {American Physical Society},
  doi = {10.1103/PhysRevLett.125.154501},
  urldate = {2023-06-19}
}

@article{page-zaki-2014,
  title = {Streak Evolution in Viscoelastic {{Couette}} Flow},
  author = {Page, Jacob and Zaki, Tamer A.},
  year = {2014},
  month = mar,
  journal = {J. Fluid Mech.},
  volume = {742},
  pages = {520--551},
  issn = {0022-1120, 1469-7645},
  doi = {10.1017/jfm.2013.686},
  urldate = {2026-08-13},
  langid = {english}
}

@article{peskin-2002,
  title = {The Immersed Boundary Method},
  author = {Peskin, Charles S.},
  year = {2002},
  journal = {Acta Numer.},
  volume = {11},
  pages = {479--517},
  publisher = {Cambridge University Press},
  issn = {1474-0508, 0962-4929},
  doi = {10.1017/S0962492902000077},
  urldate = {2021-12-30},
  langid = {english}
}

@article{poggi-etal-2004,
  title = {The {{Effect}} of {{Vegetation Density}} on {{Canopy Sub-Layer Turbulence}}},
  author = {Poggi, D. and Porporato, A. and Ridolfi, L. and Albertson, J. D. and Katul, G. G.},
  year = {2004},
  month = jun,
  journal = {Bound.-Layer Meteorol.},
  volume = {111},
  number = {3},
  pages = {565--587},
  issn = {1573-1472},
  doi = {10.1023/B:BOUN.0000016576.05621.73},
  urldate = {2023-02-01},
  langid = {english}
}

@article{rosti-2026,
  title = {Simulating Laminar and Turbulent Multiphase Flows with {{Fujin}}},
  author = {Rosti, M. E.},
  year = {2026},
  journal = {Fluid Dyn. Res.},
  volume = {58},
  pages = {021401},
  issn = {1873-7005},
  doi = {10.1088/1873-7005/ae4fee},
  urldate = {2026-03-18},
  langid = {english}
}

@article{sharma-garciamayoral-2020-1,
  title = {Scaling and Dynamics of Turbulence over Sparse Canopies},
  author = {Sharma, A. and {Garc{\'i}a-Mayoral}, R.},
  year = {2020},
  month = apr,
  journal = {J. Fluid Mech.},
  volume = {888},
  pages = {A1},
  publisher = {Cambridge University Press},
  issn = {0022-1120, 1469-7645},
  doi = {10.1017/jfm.2019.999},
  urldate = {2024-01-02},
  langid = {english}
}

@book{spagnolie-2014,
  title = {Complex {{Fluids}} in {{Biological Systems}}: {{Experiment}}, {{Theory}}, and {{Computation}}},
  shorttitle = {Complex {{Fluids}} in {{Biological Systems}}},
  author = {Spagnolie, S. E.},
  year = {2014},
  month = nov,
  publisher = {Springer},
  googlebooks = {WoqeBQAAQBAJ},
  isbn = {978-1-4939-2065-5},
  langid = {english}
}

@article{sugiyama-etal-2011,
  title = {A Full {{Eulerian}} Finite Difference Approach for Solving Fluid--Structure Coupling Problems},
  author = {Sugiyama, K. and Ii, S. and Takeuchi, S. and Takagi, S. and Matsumoto, Y.},
  year = {2011},
  month = feb,
  journal = {J. Comput. Phys.},
  volume = {230},
  number = {3},
  pages = {596--627},
  issn = {0021-9991},
  doi = {10.1016/j.jcp.2010.09.032},
  urldate = {2023-06-19},
  langid = {english}
}

@article{sundin-bagheri-2019,
  title = {Interaction between Hairy Surfaces and Turbulence for Different Surface Time Scales},
  author = {Sundin, J. and Bagheri, S.},
  year = {2019},
  month = feb,
  journal = {Journal of Fluid Mechanics},
  volume = {861},
  pages = {556--584},
  issn = {0022-1120, 1469-7645},
  doi = {10.1017/jfm.2018.935},
  urldate = {2025-10-02},
  langid = {english}
}

@article{taylor-1938,
  title = {The {{Spectrum}} of {{Turbulence}}},
  author = {Taylor, G. I.},
  year = {1938},
  journal = {Proc. R. Soc. Lond. A},
  volume = {164},
  number = {919},
  pages = {476--490},
  publisher = {Royal Society},
  doi = {10.1098/rspa.1938.0032},
  urldate = {2024-07-24}
}

\section*{Data availability statement}
All data needed to evaluate the conclusions of this work are present in the main text and/or the Supplementary Information. Data required to reproduce the figures are available on the website of the Complex Fluids and Flows Unit at OIST (\url{https://www.oist.jp/research/research-units/cffu/publications/publication-data}). Further details regarding simulations or experiments might be provided upon reasonable request to the corresponding author.

\section*{Code availability statement}
Full details on the code implementation, validation, and related resources are available on the website of the Complex Fluids and Flows Unit at OIST (\url{www.oist.jp/research/research-units/cffu/fujin}).

\section*{Author contributions}
G.F.R.~performed the numerical simulations; R.A.L.C.~performed the experiments, with support from S.J.H.; G.F.R.~analysed the data.
G.F.R. and M.E.R. outlined the manuscript content, with the first draft written by G.F.R. and input from R.A.L.C., S.J.H.~and A.Q.S.
A.Q.S. and M.E.R. supervised the research and acquired funding.
All authors agree on the submitted manuscript.

\section*{Funding}
This research was supported by the Okinawa Institute of Science and Technology Graduate University (OIST) with subsidy funding to A.Q.S. and M.E.R. from the Cabinet Office, Government of Japan. A.Q.S., S.J.H., and M.E.R.~acknowledge funding from the Japan Society for the Promotion of Science (JSPS), grant 24K00810. R.A.L.C.~also acknowledges funding from JSPS, with grant 23K19103. G.F.R. and M.E.R. acknowledge the computer time provided by the Scientific Computing and Data Analysis section of the Core Facilities at OIST, and the computational resources offered by the HPCI System Research Project with grant hp260009.

\end{document}


\preprint{APS/123-QED}

\title{\small Supplementary Information for the paper \\
\vspace{.5cm}
\normalsize Coexisting order and chaos in micro-textured elastic flows}

 \author{Giulio Foggi Rota\orcidA{}}
 \affiliation{Complex Fluids and Flows Unit, Okinawa Institute of Science and Technology Graduate University, Okinawa 904-0495, Japan}

 \author{Ricardo Arturo Lopez de la Cruz\orcidB{}}
 \affiliation{Micro/Bio/Nanofluidics Unit, Okinawa Institute of Science and Technology Graduate University, Okinawa 904-0495, Japan}
 
 \affiliation{Department of Mechanical Systems Engineering, Faculty of Engineering, Shinshu University, 4-17-1 Wakasato,  Nagano 380-8553, Japan}

 \author{\\ Simon J. Haward\orcidC{}}
 \affiliation{Micro/Bio/Nanofluidics Unit, Okinawa Institute of Science and Technology Graduate University, Okinawa 904-0495, Japan}

 \author{Amy Q. Shen\orcidD{}}
 \affiliation{Micro/Bio/Nanofluidics Unit, Okinawa Institute of Science and Technology Graduate University, Okinawa 904-0495, Japan}

 \author{Marco Edoardo Rosti\orcidE{}}
 \email{marco.rosti@oist.jp}
 \affiliation{Complex Fluids and Flows Unit, Okinawa Institute of Science and Technology Graduate University, Okinawa 904-0495, Japan}

\maketitle

\section{Governing equations}

Viscoelastic fluid flow is described by the Navier-Stokes equations for conservation of momentum, along with the incompressibility constraint for conservation of mass. Denoting the flow velocity and pressure with $\mathbf{u}$ and $p$, respectively, the governing equations write
\begin{equation}   
   \rho (\partial_t \mathbf{u} + \boldsymbol{\nabla} \cdot \mathbf{u u}) = - 
   \boldsymbol{\nabla} p + \mu_f \boldsymbol{\nabla} \cdot (\boldsymbol{\nabla} \mathbf{u} 
   + (\boldsymbol{\nabla} \mathbf{u})^T) + \boldsymbol{\nabla} \cdot \mathbf{T} + 
   \rho \mathbf{f}, \text{ and}   
   \label{eq:ns}
\end{equation} 
\begin{equation}   
   \boldsymbol{\nabla} \cdot \mathbf{u} = 0,
   \label{eq:inc}
\end{equation} 
where $\mathbf{f}$ represents any body force applied to drive the flow, and $\mathbf{T}$ is the non-Newtonian stress tensor. 
In our simulations, the fluid is driven by a dynamically adjusted body force $f\hat{\mathbf{x}}$ along the streamwise direction, maintaining a constant instantaneous domain-averaged flow speed $U$.
 
An additional transport equation describes the evolution of the polymer conformation tensor $\mathbf{C}$,
\begin{equation}   
   \partial_t{\mathbf{C}} + (\mathbf{u} \cdot \boldsymbol{\nabla}) \mathbf{C} - 
   \mathbf{C} \cdot \boldsymbol{\nabla} \mathbf{u} - (\boldsymbol{\nabla} \mathbf{u})^T 
   \cdot \mathbf{C} = - \frac{1}{\lambda} (g(tr \mathbf{C}) \mathbf{C} - \mathbf{I}), 
   \label{eq:conf}
\end{equation} 
where $\lambda$ is the intrinsic relaxation timescale of the polymeric molecules diluted in the Newtonian solvent in the absence of external forcing, and $g$ denotes the functional relation between $\mathbf{C}$ and the non-Newtonian stress $\mathbf{T} = ({\mu_p}/{\lambda})(g(tr \mathbf{C})\mathbf{C} - \mathbf{I})$. Consistently with the Oldroyd-B model, we assume that the polymer molecules stretch indefinitely and deform affinely with the flow, so that $g(tr \mathbf{C}) = 1$. 

To isolate intrinsic fluid motions from local adaptations to the canopy geometry, throughout the manuscript we decompose any field $\mathbf{\Gamma}$ as
\begin{equation}
   \mathbf{\Gamma}(\mathbf{x},t) = \overline{\mathbf{\Gamma}}(\mathbf{x}) + 
   \mathbf{\Gamma}^\prime(\mathbf{x},t),
   \label{ref:decomp}
\end{equation}
where $\overline{\mathbf{\Gamma}}$ denotes the time-average and $\mathbf{\Gamma}^\prime$ the residual fluctuations.

\section{Comparison of the experimental and numerical setups}

\begin{figure}
   \includegraphics[width=.3\textwidth]{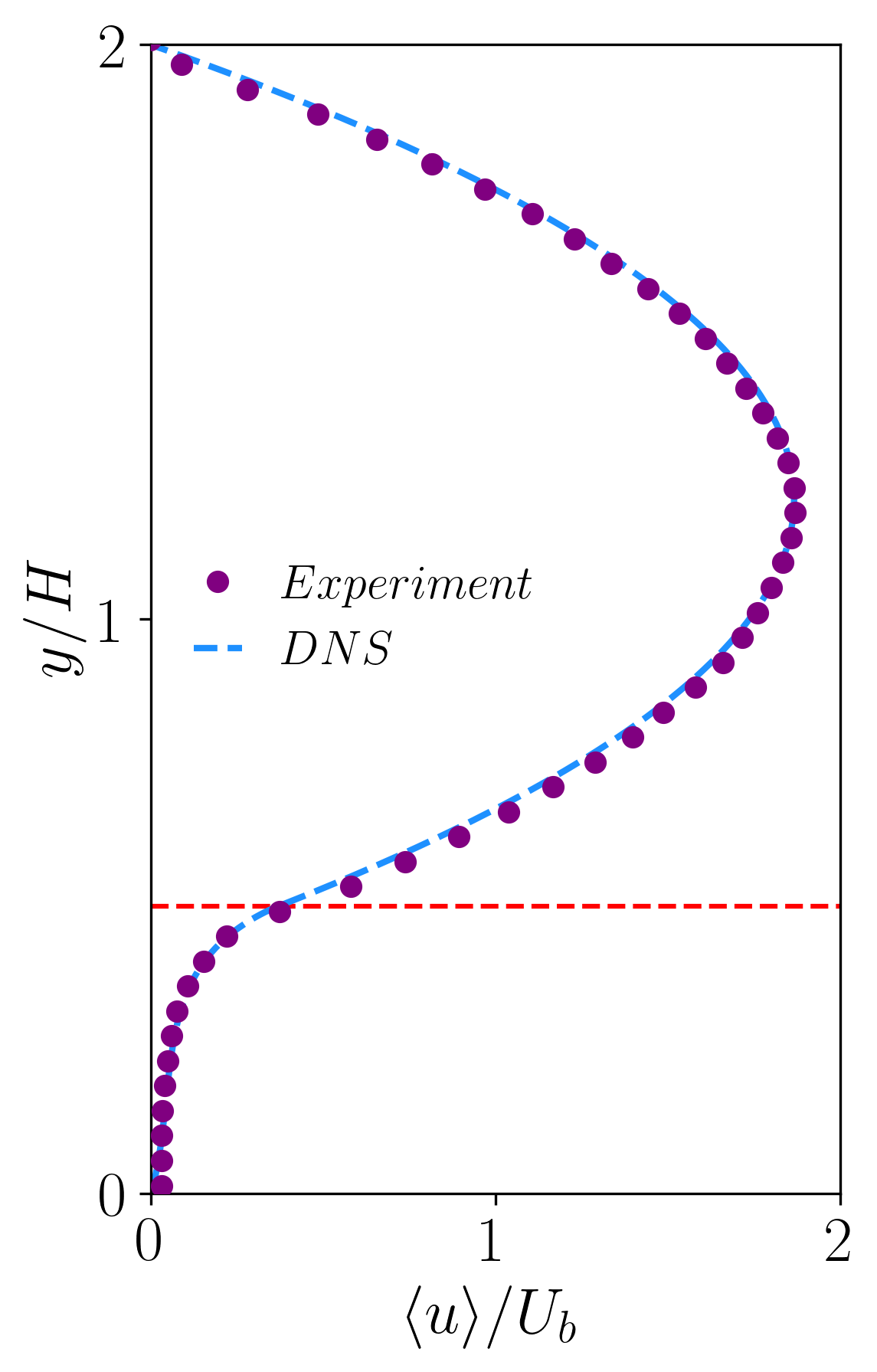}
   \hspace{2cm}
   \raisebox{1.3cm}{%
    \includegraphics[width=.37\textwidth]{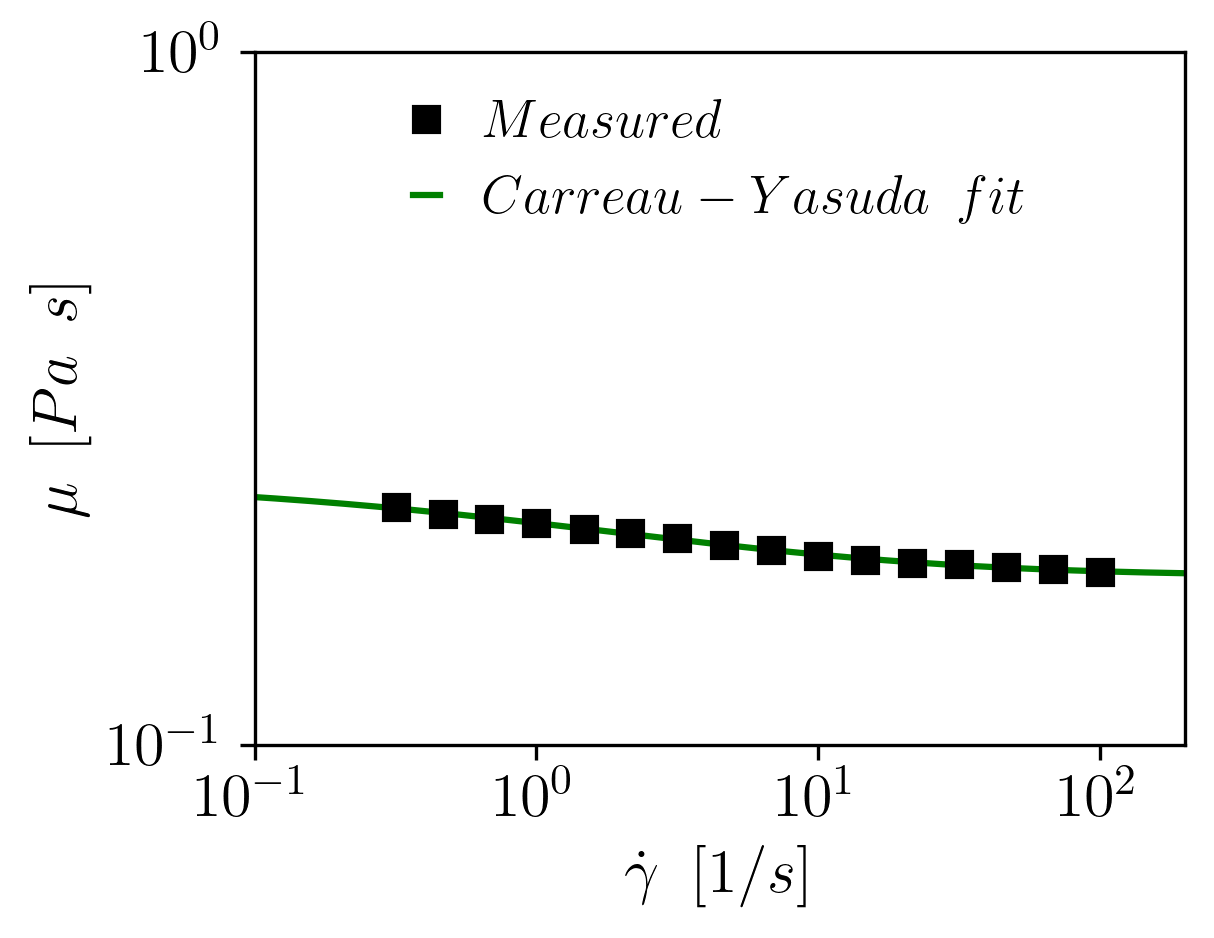}%
     }
   \caption{\textbf{Left: Comparison of laminar velocity profiles} in simulations (blue dashed line) and experiments (purple dots), for the canopy flow of a Newtonian liquid as detailed in the text. The canopy tip is indicated by a red dashed line. The good agreement confirms geometric equivalence between the experimental and computational setups, and corroborates the comparison of results presented in the main text.
   \textbf{Right: Flow curve} of the experimental test fluid (black squares) fit with a Carreau-Yasuda model (green curve).}
   \label{fig:S1}
\end{figure}

Unavoidable differences exist between the experimental and numerical setups, owing to the inherent limitations of each approach. Experiments are performed in a duct (bounded along the $z$ direction) while simulations employ a channel (periodic in $z$). Furthermore, neither the laser-etching technique nor the one-dimensional representation of the pillars via the Lagrangian immersed boundary method offers complete control over their diameter. It is therefore essential to verify that, despite these differences, the two setups are geometrically equivalent once the relevant flow parameters are matched. To this end, we compare the laminar velocity profiles of a Newtonian liquid flowing at $Re \approx 0.5$ over a canopy with solidity $\mathcal{S} \approx 0.5$ and height $h/H \approx 0.5$, measured in experiments and obtained from simulations. Remarkable agreement is found between the two, as shown in the left of figure~\ref{fig:S1}, corroborating the meaningful comparison of results across the two approaches.

The viscosity of the test fluid used in experiments is measured with a cone-plane geometry (50 mm in diameter and a 1\textdegree angle) in a stress controlled rheometer. The corresponding flow curve is shown in the right of figure~\ref{fig:S1}. The fluid is slightly shear-thinning with a change of approximatively 0.07 Pa s over about two decades of shear rate. Data is fit with a Carreau-Yasuda model with the following equation:
\begin{equation}
    (\mu-\mu_\infty )/(\mu_0-\mu_\infty )=[1+(\dot{\gamma}/\dot{\gamma}^*)^a ]^{((n-1)/a)}.
\end{equation}
Here, $\mu_0=0.243 \enspace Pa \enspace s$ and $\mu_\infty=0.175 \enspace Pa \enspace s$ represent the zero shear and infinite shear viscosities, respectively. The applied shear rate is denoted by $\dot{\gamma}$, while
$\dot{\gamma}^*=6.15 \enspace 1/s$ is the characteristics threshold for the onset of shear-thinning effects. The power-law exponent is $n=0.011$, and $a=0.487$ is a fitting parameter. As a reference, the average velocity profile at low $De$ does not show the typical flat profile of a shear-thinning fluid, as observed in other studies with stronger shear-thinning behaviour \cite{lopez-haward-shen-2025}. Comparison with the Oldroyd-B fluid used in simulations thus appears well motivated and supported.

\begin{figure}
    \centering
    \includegraphics[width=.34\textwidth]{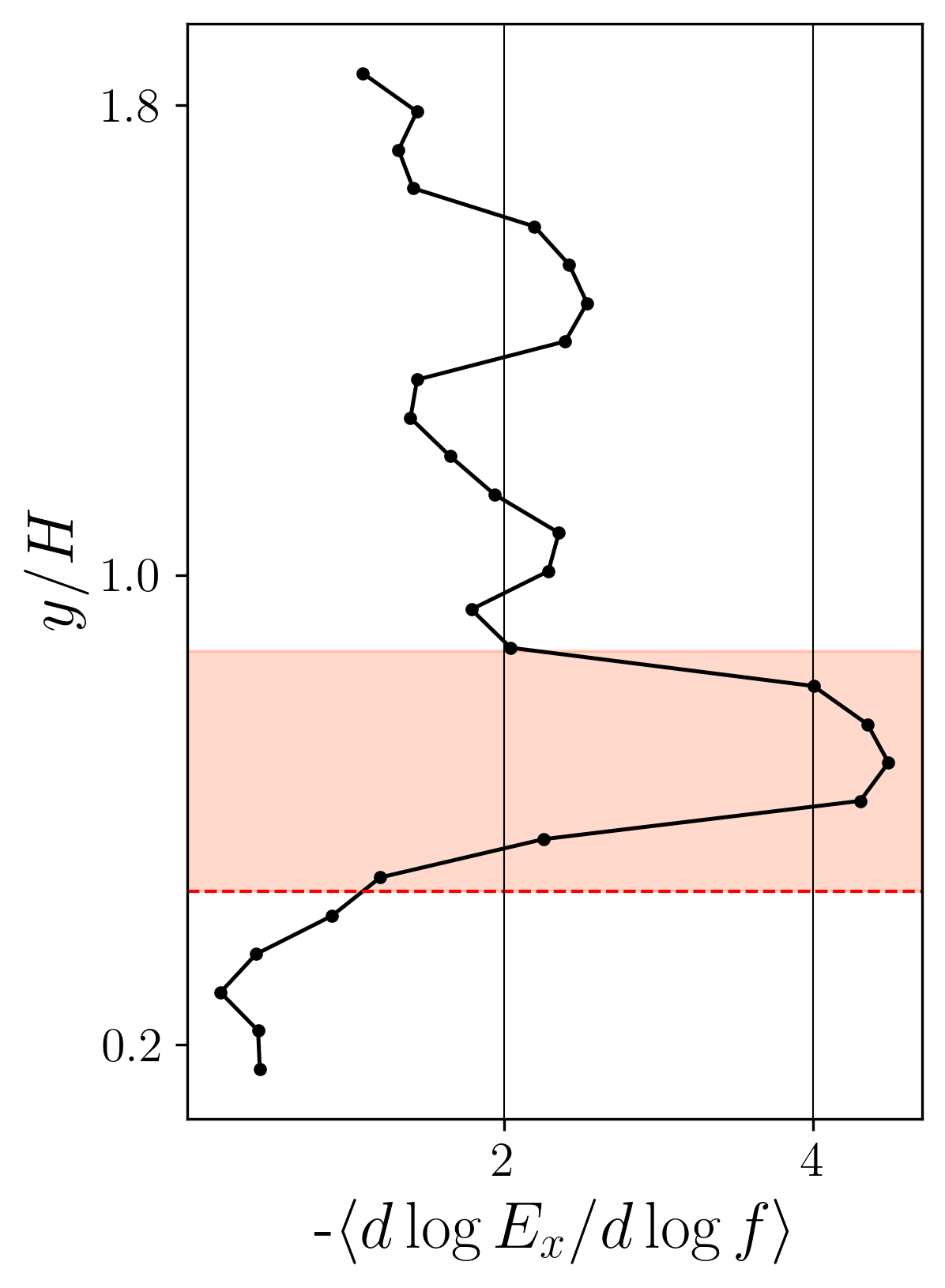}
    \caption{\textbf{Logarithmic derivative of the temporal spectra measured in our experiments.}  The spectra appear to decay as $f^{-4}$ within the elastically active layer, while above that they recover the $f^{-2}$ trend conventional for channel flows.}
    \label{fig:S2}
\end{figure}

As noted in the main text, our simulations show that the coherent pattern formed within the elastically active layer advects downstream at the mean speed of the layer itself, $U_l$. However, within our computational domain ($4\pi H \approx 12.6 H$) the pattern appears space-filling (figure~3a of the main text), so its characteristic streamwise extent cannot be measured directly; probing a substantially larger domain is however beyond current computational feasibility. 
Experiments offer a complementary view. The channel is long enough to host a much greater streamwise extent than what accessible in the simulations, but only a narrow window of about $10H$ is imaged, again preventing a direct spatial measurement of the pattern. The fixed imaging window does, however, allow long observation times, giving access to the temporal autocorrelation of the streamwise velocity fluctuations (inset of figure~3a in the main text). This first crosses zero at $\Delta_t \sim 13 H/U$, suggesting that one period of the underlying pattern (four times the first zero-crossing) spans $\Delta_t \sim 50 H/U$ (compatibly with figure~1e of the main text).
Using the mean advection velocity of the layer ($U_l$) to convert the experimentally observed period into a spatial wavelength, we obtain a characteristic streamwise wavelength $l_x\sim 24 H$. This length exceeds both the experimental imaging window and the simulation domain, confirming that the pattern is genuinely space-filling in our simulations and unobservable in a single experimental frame -- consistent with why it is only recovered here through a temporal reconstruction. Finally, this reasoning yields the conversion factor for recasting the spatial correlation $\mathcal{R}_{u^\prime u^\prime,x}$ from simulations in a function of time $\mathcal{R}_{u^\prime u^\prime,t}$, in the inset of figure~3a from the main text. Accounting for the pattern compression induced by the simulation domain, $\Delta_x=U_l (L_x/l_x) \Delta_t \sim 0.2  \Delta_t$, where $L_x=4\pi$ is the streamwise extent of the simulation domain.

The peculiar yo-yoing behaviour of the phase-averaged velocity $\langle u \rangle$ over the canopy tip is isolated upon rescaling it and the gradient coordinate in the standard fashion of mixing layers \cite{ghisalberti-nepf-2002}.
At every position $(\phi_x,z)$, a single profile $\langle u \rangle\rvert_{\phi_x,z}(y)$ is attained. We thus denote $U_1$ the average profile value within the canopy array, $U_2$ the first profile maximum over the canopy tip, and $\Delta U = U_2 - U_1$ their difference. Their arithmetic mean is instead $\overline{U}=( U_1 + U_2)/2$, attained where $\langle u \rangle\rvert_{\phi_x,z}(\overline{y})=\overline{U}$.
The desired  rescaling of the gradient direction is thus retrieved by computing the momentum thickness $\theta_m=\displaystyle \int_{-\infty}^{+\infty} \left[ \frac{1}{4} - \left( \frac{\langle u \rangle\rvert_{\phi_x,z} (y)- \overline{U}}{\Delta U} \right) \right] dy$. 
Shifting everything to the centre of the mixing layer, in figure~3c of the main text we plot $(y-\overline{y})/\theta_m$ versus $(\langle u \rangle - \overline{U})/\Delta U$ for different choices of $\phi_x$, with $z$ fixed at the domain centre.

\section{Further particulars on the analysis of experimental data}
To compute the temporal spectra reported in the bottom inset of figure~2a from the main text, we sample the experimental field of the streamwise velocity fluctuations with frequency $0.7 U/H$, and instantaneously average it along the streamwise and spanwise directions. By doing so we are able to reduce noise, and isolate the temporal dynamics of the flow within the PIV observation window at different positions above the canopy. As just noted above, the frame is in fact too short to capture any appreciable streamwise variation. We then window and Fourier-transform the velocity signal preserving its $y$ dependence, eventually attaining the temporal spectrum of the kinetic energy contained in the streamwise velocity fluctuations as a function of the vertical $y$ coordinate and of the frequency $f$. Computing its average logarithmic derivative $-\langle d \log E_x / d \log f \rangle$ at frequencies higher than $0.04 U/H$ (in figure~\ref{fig:S2}), the average slope of the scaling range reaches values close to $4$ within a range consistent with the elastically active layer described in the main text, settling to $2$ above it. The heights sampled in the main text are thus chosen to represent these two regions.

We then shift our attention to figure~2b of the main text, where we demonstrate the correspondence between simulated and experimental shear balance components. Only the Reynolds and viscous stresses are directly available from experimental measurements. We therefore assume a linear trend of the total shear and, reasonably, the same canopy drag measured in simulations (based on the good comparison reported in the left of figure~\ref{fig:S1}). 
The viscoelastic shear stress of the experiments, then, immediately follows by closure.

\pagebreak
\renewcommand{\bibsection}{{ \section*{References}}}
\bibliographystyle{naturemag}
\bibliography{Wallturb}